\documentstyle[twoside,fleqn,psfig,amssymb]{actaps}
\begin{document}\headings{1}{7}
\title{GENERATION AND MEASUREMENT OF NONCLASSICAL STATES BY 
QUANTUM FOCK FILTER}\author{G. M. D'Ariano, L. Maccone, M. G. A. Paris, 
and M. F. Sacchi}{{\sc Theoretical Quantum Optics Group} -- INFM Unit\'a 
di Pavia \\ Dipartimento di Fisica 'Alessandro Volta' -- Universit\'a di 
Pavia \\ via A. Bassi 6,  I-27100 Pavia, Italy}\def\authorlist{G. M. D'Ariano, 
L. Maccone, M. G. A. Paris and M. F. Sacchi}\def\shorttitle{Generation and
measurement by quantum Fock filter}
\abstract{We study a novel optical setup which is able to select a 
specific Fock component from a generic input state. The device allows 
to synthesize number states and superpositions of few number states, 
and to measure the photon distribution and the density matrix 
of a generic input signal.}  
\section{Introduction}
In the last two decades in Quantum Optics there has been a fast and 
exciting development. Many experiments involving the optical field  
fall into the quantum domain, and many quantum mechanical {\em gedanken}
experiments are now performed in optical laboratories. Basic concepts of 
quantum mechanics such as the reduction postulate, the uncertainty relation, 
the Sch\"odinger cat and nonlocal phenomena play a major role and their 
effects can be directly observed. 
Besides this fundamental interest, the field receives attention 
also in view of future applications, which are mainly motivated by the 
potential improvement offered by Quantum Mechanics to the manipulation 
and the transmission of information. Such recent developments have renewed 
the interest for two fundamental themes in Quantum Optics, namely the 
{\it generation} and the {\it measurement} of nonclassical 
states of the radiation field. \par
In this paper, we address both these aspects and suggest a novel all-optical 
device which is able to select a fixed Fock component from a traveling wave 
initially prepared in a generic (possibly mixed) state. The scheme, which 
we named {\em Fock Filter}, consists of a ring cavity coupled to the signal 
through a cross-Kerr medium. At the output of the device, the signal and the 
cavity modes are strongly entangled, so that a successful photodetection 
of the cavity modes reduces the signal into a state with the desired number 
of photons. The proposed setup has two main applications. On one hand, it 
can be used to synthesize a generic number state $|n\rangle$ or a 
superposition of few number states, say $|\psi\rangle\propto\alpha |n_1
\rangle+ \beta |n_2\rangle $, starting from a coherent source. On the other 
hand, it allows to measure the photon distribution $P(n)=\langle n|\hat\nu 
|n\rangle$ and the density matrix $\hat\nu_{nk}=\langle n|\hat\nu |k\rangle$ 
of a generic input signal $\hat\nu$.
\section{The Fock Filter}
The device we propose is schematically depicted in Fig. 1. It consists of an 
active ring cavity coupled to the signal by a cross-Kerr 
medium. The cavity has a high quality factor, namely it is built by low
transmissivity (denoted by $\tau$) beam splitters. The cross-Kerr interaction
couples the cavity mode $d$ to the traveling signal mode $c_1$, according to 
the  unitary evolution  $\hat U_{K}=\exp(-i\chi t d^\dag d c_1^\dag c_1)$.
In this way, the cavity mode experiences a phase-shift which depends on the
quantum state of the signal mode. A further tunable phase-shift $\psi$ is also 
inserted in the cavity path. A port of the cavity (mode $a_1$) is fed by a 
strong coherent probe, whereas the second port (mode $a_2$) is left vacuum. 
At the output of the cavity, the mode $b_1$ is simply absorbed, whereas the 
mode $b_2$ is monitored by an avalanche photodetector. As we will see in the 
following, we only need to know whether or not any photon is present, namely to 
perform an {\sf ON/OFF} photodetection. 
\begin{figure}[h]
\begin{minipage}{70mm}
\psfig{file=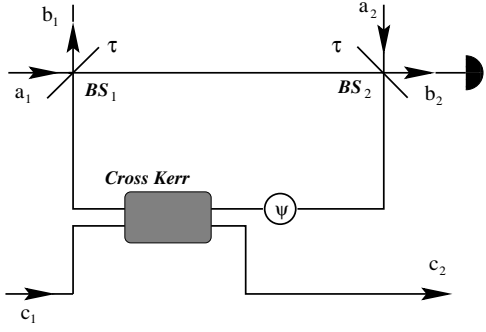,width=64mm}
\end{minipage}\hspace{8pt}\begin{minipage}{65mm}
\caption{Fig.1: Schematic diagram of the Fock Filter. BS$_1$ and 
BS$_2$ denote low transmissivity beam splitters, and $\psi$ a tunable phase 
shift. The cavity input modes $a_1$ and $a_2$ are in a coherent state 
and in the vacuum respectively. The box denotes the cross Kerr medium that 
couples the cavity mode with the signal mode. A successful detection at the 
photodiode D filters the desired Fock component from the input signal.
\label{f:stp}}\end{minipage}
\end{figure}\par
The probability operator measure (POM)
describing such a detection scheme is given by the the two-value operator
$\hat\Pi_0\doteq\sum_{k=0}^\infty(1-\eta)^k|k\rangle\langle
k|$ and $\hat\Pi_1\doteq \widehat{{\Bbb I}} -\hat\Pi_0$
where $\eta$ is the quantum efficiency of the photodetector. 
The modes transformation of the cavity is expressed by [1]
\begin{eqnarray}
\left\{\begin{array}{l}b_1=\kappa(\varphi)
a_1+e^{i\varphi}\sigma(\varphi) a_2\\ b_2=\sigma(\varphi) 
a_1+\kappa(\varphi) a_2\end{array}\right.
\;,\label{ring}
\end{eqnarray}
where the phase-dependent transmissivity $\sigma$ and
reflectivity $\kappa$ amplitude are given by
\begin{eqnarray}
\kappa(\varphi)\doteq\frac{\root\of{1-\tau}(e^{i\varphi}-1)}
{1-[1-\tau]e^{i\varphi}}\qquad \sigma(\varphi)\doteq\frac\tau{1-
[1-\tau]e^{i\varphi}},\label{defsigma}
\end{eqnarray}
with $\left|\kappa(\varphi)\right|^2+\left|\sigma(\varphi)\right|^2
=1$. The phase $\varphi$ is the total phase-shift experienced by the
cavity mode, namely the sum of the shift due to the Kerr interaction 
and the tunable shift $\psi$. For the signal in a number state $|n\rangle$ 
the total phase shift imposed to the cavity mode is $\varphi\equiv \varphi_n
=\psi - \chi n t$. In order to simplify notation, we write $\sigma _n\doteq 
\sigma(\varphi _n)$ and analogously $\kappa_n \doteq \kappa (\varphi_n)$. 
The input state of the device can be written as 
$ \hat\varrho_{in}=|\alpha\rangle\langle\alpha|\otimes|0\rangle\langle 0|
\otimes\hat\nu_{in}$, 
namely a generic state $\hat\nu_{in}$ for the signal mode and a
strong coherent state $|\alpha\rangle$ for the probe, the second port 
of  the cavity being left unexcited. The output state can be easily found
in the Schr\"odinger picture as 
\begin{eqnarray}
\hat\varrho_{out} = \label{rhoout} 
\sum_{n,m=0}^{\infty} \nu_{nm} \: |\kappa_n\alpha\rangle
\langle \kappa_m\alpha|\otimes|\sigma_n\alpha\rangle\langle
\sigma_m\alpha|\otimes |n\rangle\langle m|\;.
\end{eqnarray} 
The measurement scheme consists in detecting whether (detector D on) 
or not (detector D off) any photon is present at the output of the cavity.
The corresponding probabilities are given by
\begin{eqnarray} 
P_1={\mbox Tr}[\hat\Pi_1\hat\varrho_{out}] = \sum_{n=0}^\infty \nu_{nn} 
\left(1- e^{-\eta|\alpha|^2\:|\sigma_n|^2}\right) \qquad 
P_0 = 1-P_1 \label{P1} \;,
\end{eqnarray} 
whereas the output signal conditioned by the photodetection reads 
\begin{eqnarray} 
\hat\nu_{out} ({\sf ON}) &=& \frac{e^{-|\alpha|^2}}{P_1}
\sum_{n,m=0}^\infty \nu_{nm}\: e^{|\alpha|^2 \big[\kappa_n\kappa^*_m +
\sigma_n\sigma^*_m \big]}\left(1- e^{-\eta|\alpha|^2 \sigma_n\sigma^*_m}
\right)\:|n\rangle\langle  m| \label{nuon}
\end{eqnarray}
The filtering properties of the device are due to the strong dependence of 
the cavity transmissivity on the internal phase-shift. The overall 
transmissivity function writes as
\begin{eqnarray}
|\sigma_n|^2 = \left[1+ 4 \frac{1-\tau}{\tau^2}\sin^2\frac{\psi-\chi nt}{2} 
\right]^{-1}\label{sigma}\;,
\end{eqnarray}
and exhibits a periodic structure sharply peaked at 
$n = n^*+ 2\pi j / (\chi t)$ with $n^*=\psi/(\chi t)$ and 
$j\in{\mathbb Z}$. In the peaks, it has unit height and width of the order 
of the beam splitter transmissivity $\tau$ (typically $\tau\sim 1\% - 0.01\%$).
The value $n^*$ can be adjusted to an arbitrary integer by tuning the 
phase-shift $\psi$ as a multiple of $\chi t$, whereas multiple resonances 
are avoided by using relatively small values of the nonlinearity $\chi t$, so 
that the values of $n$ for $j\not = 0$ correspond to 
vanishing matrix elements $\nu_{ni}\simeq 0\ \forall i$. In this way the 
cavity is set at resonance only by a single Fock component $|n^*\rangle$ 
of the signal, which is {\em filtered} at the output in the case of 
successful photodetection. In the next section we will analyze this process 
in more detail.
\section{Synthesis of number states}
Let us now consider a cavity with a high quality factor (i.e. $\tau \ll 1$) 
adjusted to select the Fock component $n^*$ by tuning $\psi=\chi t n^*$. 
In this case, the detection probability and the conditional output states of 
Eqs. (\ref{P1}) and (\ref{nuon}) rewrite as follows 
\begin{eqnarray}
P_1 \simeq \nu_{n^*n^*}+\frac{\eta |\alpha |^2 \tau^2}{(\chi t)^2}\sum_{p\neq 
n^*}\frac{\nu_{pp}}{(n^* -p)^2} \;,\label{P1asym}
\end{eqnarray} and
\begin{eqnarray}
\hat\nu_{out} ({\sf ON}) \simeq \frac{1}{\sqrt{\cal N}}\left[ |n^*\rangle
\langle n^*|+\frac{\eta |\alpha|^2 \tau^2}{(\chi t)^2}\sum_{n,k\neq n^*}
\frac{\nu_{nk}}{(n^*-n)(n^*-k)}|n\rangle\langle k|\right]\;,\label{nuonasym}
\end{eqnarray}
where ${\cal N} = 1 + (\eta |\alpha|^2 \tau^2)/(\chi t)^2 \sum_{p\neq n^*}
\nu_{pp}/(p-n^*)^2$ is a normalization constant. Both equations are valid when small values of the nonlinearity are involved, namely when a single Fock 
component sets the cavity into resonance. The physical meaning of Eqs. 
(\ref{P1asym}) and (\ref{nuonasym}) is apparent: when the cavity is "good
enough" to appreciate the phase-shift imposed by the passage of the desired
Fock component [i.e. when $\tau \ll (\chi t)$] the detection probability
equals the probability of having $n^*$ photons in the signal $P_1 \simeq 
\nu_{n^*n^*}$, and the conditional output state approaches the corresponding 
number state $\hat\nu_{out} ({\sf ON}) \simeq |n^*\rangle\langle n^*|$, which 
is {\em synthesized} from the input signal. Eqs. (\ref{P1asym}) and 
(\ref{nuonasym}) also illustrate the effect of nonunit quantum efficiency of 
the probe photodetector. If $\eta$ is lower than 
$100\%$, the detection probability decreases, and thus also the synthesizing
rate. However, the synthesized state is closer to the desired number state, 
namely the synthesizing quality is improved.  In Fig. 2 the synthesis 
of the number state $|n^* \equiv 4\rangle$ is illustrated for decreasing values
of the beam splitter transmissivity. 
\begin{figure}[h]
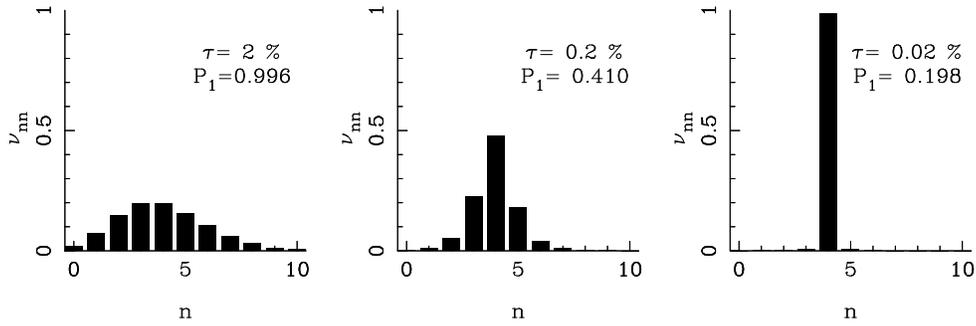

\begin{tabular}{ccc}
\psfig{file=num1.ps,width=4cm}&\psfig{file=num2.ps,width=4cm}&
\psfig{file=num3.ps,width=4cm}\end{tabular}
\caption{Fig.2: The photon distribution of the conditional output state $\hat 
\nu_{out} ({\sf ON})$ for different values of the transmissivity $\tau$, 
reported on each plot. The plots refer to the synthesis of the  number state 
$|n^* \equiv 4 \rangle$, with $\chi t = 0.01$, $\psi=0.04$ and $\eta=0.8$. Both
the probe mode $a_1$ and the signal mode $c_1$ are excited in a coherent state,
with real amplitude $\alpha =20$ and $\beta =2$ respectively. The probabilities
of obtaining the three states (i.e. the detection probability at D) are also 
indicated on each plot. The last plot shows that a transmissivity $\tau=0.02\%$
is sufficient for a reliable synthesis of the desired number state.} 
\end{figure}
\par The Fock Filter behaves differently when larger nonlinearities or quite 
excited input signals are involved. In this case, the cavity may be set at 
resonance by several Fock components of the signal mode, corresponding to 
different integers that are multiple of $\chi t$. Let us consider, as an 
example, the situation in which two Fock components, corresponding to the 
values $n_1\equiv n^*$ and $n_2=n_1+2\pi/(\chi t)$, set the cavity into 
resonance. For $\tau \ll \chi t$ Eqs. (\ref{P1}) and (\ref{nuon}) can be 
written as $P_1 \simeq \nu_{n_1n_1}+ \nu_{n_2n_2}$ and
$\hat\nu_{out}({\sf ON})\simeq 1/P_1 \Big[\nu_{n_1 n_1}|n_1\rangle\langle  
n_1|+\nu_{n_2 n_2}|n_2 \rangle\langle n_2| + \nu_{n_1 n_2}|n_1 \rangle\langle  
n_2|+\nu_{n_2 n_1}|n_2 \rangle\langle  n_1|\Big]$.
If the input signal $\hat\nu$ is excited in a coherent state, one has
$\nu_{n_1 n_1}\nu_{n_2 n_2}=\nu_{n_1 n_2}\nu_{n_2 n_1}$, and hence 
$\hat\nu_{out}({\sf ON})$ is a pure state. Actually, by varying the amplitude 
of the input coherent signal any superposition of the form $|\psi\rangle
\propto \alpha |n_1\rangle+ \beta |n_2\rangle$ may be synthesized at the 
output. We just mention that this kind of superposition is the paradigm for 
the realization of an optical qubit.
\section{State measurement}
In this section we show how the Fock filter can be used to measure the photon
distribution, and the whole density matrix, of a generic input signal. The
method is based on Eq. (\ref{P1}), which shows that for moderate nonlinearities
and low beam splitter transmissivity ($\tau \ll \chi t < 1$) the
detection probability $P_1$ at the output of the cavity is equal to the 
diagonal matrix element of the signal $P_1\simeq\nu_{n^*n^*}$ corresponding 
to the 
integer $n^*$ selected by tuning the phase $\psi$. Therefore, by repeated
preparations of the signal and by varying the cavity tuning $\psi = \chi t 
n^*$, $n^*=0,1,2,...$ in order to span the whole excitation spectrum of the 
signal it is possible to record the photon number distribution of a generic 
input state. Actually, such a measurement may be implemented by a more 
efficient scheme using a set of cavities in cascade, each tuned on a different 
integer $n_k$. 
\begin{figure}[h]
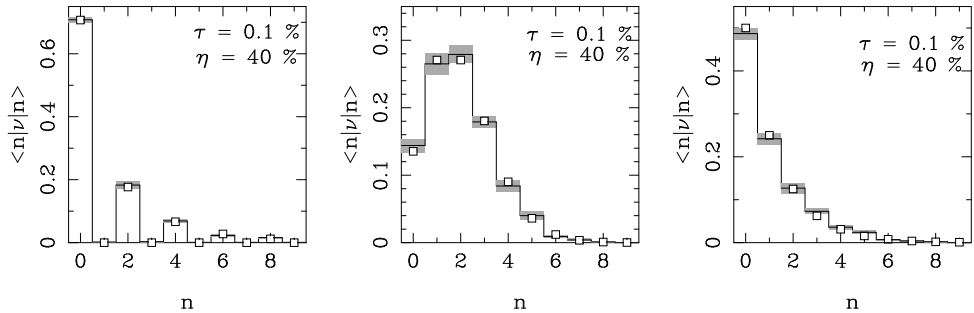

\begin{tabular}{ccc}
\psfig{file=pdn1.ps,width=4cm}&\psfig{file=pdn2.ps,width=4cm}&
\psfig{file=pdn3.ps,width=4cm}\end{tabular}
\caption{Fig.3: Monte Carlo simulations of the detection of the photon number
distribution by the Fock filter with $\chi t=0.1$, $\tau =0.1 \%$ and $\eta =
40 \%$. The distributions for a squeezed vacuum with $\langle a^\dag a\rangle
=1$ average photons, a coherent state with $\langle a^\dag a\rangle=2$ average 
photons and a thermal state with $\langle a^\dag a\rangle=1$ average photons 
are reported from the left to the right together with the corresponding
confidence interval. The empty squares indicates the theoretical values. 
In all cases a sample of $2000$ data has been used. } 
\end{figure}
\par
The input state of the $k$-th cavity is the output state
from the $(k-1)$-th one and this allows to largely reduce the number of
repeated preparations of the signal. 
At the $k$-th step, in the limit of good cavities, the 
detection probabilities approaches 
$P_1^{(k)} \simeq \nu_{n_kn_k}^{(k)}$ and $P_0^{(k)} \simeq 
1- \nu_{n_kn_k}^{(k)}$
where the density matrix $\hat\nu^{(k)}$ has been reduced according to 
the result of the detection at the $(k-1)$-th photodiode as in 
Eq.(\ref{nuon}). For good cavities one has  
\begin{eqnarray}
\hat\nu_{out}^{(k)}({\sf ON}) \simeq |n_k\rangle\langle n_k|\;, \qquad 
\hat\nu_{out}^{(k)}({\sf OFF})\simeq \sum_{p\neq n_k}\frac{\nu_{pp}^{(k)}}
{P_0^{(k)}} |p\rangle\langle p| \label{nuth}\;.
\end{eqnarray}
We checked the whole detection strategy with a Monte Carlo simulation.
In Fig. 3 we show the photon distributions obtained for a squeezed vacuum, 
a coherent state and a thermal state at the input. Remarkably, the photon 
distributions are reliably determined using a relatively small sample of 
data and a low value for the quantum efficiency of the photodetectors. \par
The Fock filter, in conjunction with the unbalanced homodyning technique 
[2-3], allows also to measure the entire density matrix of the input signal.
In order to achieve this goal, it is necessary to mix the input signal 
$\hat\nu$ with a strong coherent state $|z\rangle$ by using a high 
transmissivity beam splitter before the set of cavities (see Fig. 4). 
In the limit $(1-\tau) \ll 1$ and $|z| \gg 1$ the state entering the set 
of cavities is the displaced
signal $\hat\nu_\gamma =\hat D(\gamma)\hat\nu\hat D^\dag (\gamma)$, where
$\hat D (\gamma)=\exp(\gamma a^\dag - \hat\gamma a)$ is the displacement
operator and $\gamma = z\sqrt{1-\tau}$. In this case, the Fock filter provides
the photon distribution of the displaced state $P_\gamma (n)=\langle n|\hat
\nu_\gamma|n\rangle$, which can be expressed in terms of the density matrix
of the original signal $\hat\nu$ by the linear relation
$ P_\gamma(n)= \sum_{km} \langle k|\hat\nu|m\rangle A_{kmn}(\gamma )$
where $A_{kmn}(\gamma )=\langle n|\hat D(\gamma)|k\rangle\langle m|\hat 
D^\dag (\gamma)| n\rangle$. By measuring the photon distribution for different
values of the displacing amplitude $\gamma$, this relation can be
inverted, leading to the reconstruction of the signal density matrix. 
\begin{figure}[h]\begin{minipage}{83mm}
\psfig{file=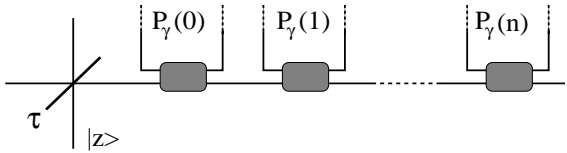,width=79mm}\end{minipage}\begin{minipage}{50mm}
\caption{Fig.4: Schematic diagram of the setup for measuring the density 
matrix with the Fock filter. The input state is mixed with a strong coherent 
state by using a high transmissivity beam splitter, and then enters the set of 
cavities in cascade, each tuned on a different Fock component.}\end{minipage}
\end{figure}
\par In particular, Opatrny et al [4] have shown that it is enough to 
measure $P_\gamma (n)$ for a fixed value of the modulus $|\gamma |$ and 
different values of the phase. In this case one has 
\begin{eqnarray}
P_{|\gamma|}^{(s)} (n) = \sum_{m=0}^{M-s} A_{m+s,m,n}(|\gamma |)
\:\nu_{m+s,m}
\label{tom1}\;,
\end{eqnarray}
where $M$ is the maximum Fock component excited in the signal, 
and $P_{|\gamma|}^{(s)} (n)$ is the Fourier transform of the photon
distribution obtained by varying the phase $\varphi =\arg (\gamma$)
The linear system (\ref{tom1}) is overdetermined and may be solved
by least squares method. The solution represents the best estimate for the 
density matrix of the input signal.
\section{Conclusions}
In this paper, we have studied a novel all-optical device: the
Fock filter, which is able to select the desired Fock component starting from 
a generic input state. The device may be used to synthesize 
number states and superpositions of few number states, as well as for 
measuring the photon distribution and the density matrix of a generic input
signal. The feasibility of the proposed setup relies on the realization
of cavities with a high quality factor, namely cavities built with low 
transmissivity beam splitters. Monte Carlo simulations have shown that
transmissivities of the order of $\tau \sim 1\% - 0.01 \%$ are needed, 
which corresponds to beam splitters currently available in optical labs.
We conclude by pointing out that the quantum efficiency of the photodetectors
is not a crucial parameter for the Fock filter: low quantum efficiency does not
degrade the performances of the device, for both the generation and the 
measurement scheme. 
\section*{Acknowledgments}
This work is part of the INFM project PRA-CAT-97 and the MURST
"Cofinanziamento 1997". MGAP thanks {\em Accademia Nazionale dei Lincei} 
for partial support through the {\em Giuseppe Borgia} award. 
\section*{References}
\begin{description}
\item{[1]} G. M. D'Ariano, L. Maccone, M. G. A. Paris,and M. F. Sacchi, 
unpublished
\item{[2]} \refer{S. Wallentowitz, W. Vogel}{Phys. Rev. A}{53}{1996}{4528}
\item{[3]} \refer{K. Banaszek, K. W\`odkievicz}{Phys. Rev.
Lett}{76}{1996}{4344}
\item{[4]} \refer{T. Opatrn\`y, D.-G. Welsch }{Phys. Rev. A}
{55}{1997}{1462}
\end{description}
\end{document}